\documentstyle[preprint,aps]{revtex}

\begin{document}

\hoffset-1cm

\draft

\preprint{TPR-96-04}

\title{Lifetimes and Sizes from Two-Particle Correlation Functions}

\author{U. Heinz, B. Tom\'a\v{s}ik, U.A. Wiedemann and  
Wu Y.-F.\cite{address}}

\address{
   Institut f\"ur Theoretische Physik, Universit\"at Regensburg,\\
   D-93040 Regensburg, Germany
}

\date{\today}

\maketitle

\begin{abstract}
We discuss the Yano-Koonin-Podgoretskii (YKP) parametrization of the 
two-particle correlation function for azimuthally symmetric expanding 
sources. We derive model-independent expressions for the YKP fit 
parameters and discuss their physical interpretation. We use them to 
evaluate the YKP fit parameters and their momentum dependence for a 
simple model for the emission function and propose new strategies for
extracting the source lifetime. Longitudinal expansion of the source 
can be seen directly in the rapidity dependence of the Yano-Koonin 
velocity.
\end{abstract}


\narrowtext

The two-particle correlation functions $C({\bf p_1}, {\bf p_2})$ of 
identical particles provide direct access to the spatio-temporal 
evolution of the collision region in heavy-ion collisions. This 
follows from the basic relation \cite{S73,P84,PCZ90,CH94} between the 
correlation function $C$ and the emission function $S(x,K)$ (here 
written down for bosons)
  \begin{equation}
     C({\bf K},{\bf q})
      \approx  1 + {\left\vert \int d^4x\, S(x,K)\,
         e^{iq{\cdot}x}\right\vert^2 \over
         \left\vert \int d^4x\, S(x,K)\right\vert^2 }\, .
  \label{1}
  \end{equation}
The emission function $S$ describes the phase space (Wigner) density 
of the boson emitting sources, and $q = p_1 - p_2$, $K = (p_1 + 
p_2)/2$ (with $p_1,p_2$ being on-shell such that $K\cdot q = 0$) 
correspond to the relative and average 4-momenta of the boson pair.  
Eq.~(\ref{1}) neglects final state interactions; for a recent review 
of methods to include the latter see Ref.~\cite{P95}. All other 
approximations leading to (\ref{1}) are well-controlled 
\cite{PCZ90,CSH95b}.  

It is the aim of Hanbury-Brown/Twiss (HBT) interferometry to extract 
information about the space-time characteristics of $S(x,K)$ from the 
measured two-particle momentum correlations by ``inverting" 
Eq.~(\ref{1}). Unfortunately, due to the on-shell conditions for the 
individual particle momenta $p_1,p_2$, this is not possible in a 
completely model-independent way: the time component $q^0$ of the 
relative momentum is constrained by 
 \begin{equation}
 \label{2}
   q^0 = \bbox{\beta}{\cdot}{\bf q}\, , \qquad
   \bbox{\beta} = {{\bf K}\over K^0} = {2{\bf K} \over (E_1+E_2)}
           \approx {{\bf K}\over E_K} 
 \end{equation} 
(with $E_K{=}\sqrt{m^2+{\bf K}^2}$ in the last step \cite{CSH95b}), 
and thus the inverse Fourier transform of Eq.~(\ref{1}) is not 
unique.  

In practice the analysis of HBT correlation data must therefore be 
based on a comparison with specific models for the source function 
$S(x,K)$, with the aim of eliminating ``unreasonable" models and 
fitting certain essential parameters (geometric size, freeze-out 
temperature, collective flow velocity, time duration of the particle 
emission process) in a class of ``reasonable" model sources. This 
procedure is enormously simplified by using so-called 
``model-independent" expressions for the HBT parameters 
\cite{CSH95b,HB95,CSH95a} which allow to calculate from an arbitrary 
source function $S$ the characteristic parameters of the two-particle 
correlation function $C$ by simple quadrature. 

While it is obvious from its definition that the single-particle 
spectrum is nothing but the zeroth space-time moment of the emission 
function, 
 \begin{equation}
 \label{5}
  E_K {dN \over d^3K} = \int d^4x \, S(x,K) \, ,
 \end{equation} 
it can also be shown \cite{CSH95b,CSH95a,WSH96} that the two-particle 
correlation function is essentially determined by its (normalized) 
second order space-time moments. Specifically, to compute the 
correlation function $C$ it is sufficient to approximate the source 
function $S$ by a Gaussian which contains only information on its 
space-time moments up to second order. To see this we write 
 \widetext
 \begin{equation}
 \label{7}
   S(x,K) = N({\bf K})\, S(\bar x({\bf K}),K)\, 
            \exp\left[ - {1\over 2} \tilde x^\mu({\bf K})\, 
            B_{\mu\nu}({\bf K})\,\tilde x^\nu({\bf K})\right]
   + \delta S(x,K) \, ,
 \end{equation} 
 \narrowtext
where
 \begin{equation}
 \label{8}
  \bar x^\mu({\bf K}) = \langle x^\mu \rangle , \ \
  \tilde x^\mu ({\bf K}) = x^\mu - \bar x^\mu({\bf K}) , \ \
  (B^{-1})_{\mu\nu}({\bf K}) 
  = \langle \tilde x_\mu \tilde x_\nu \rangle ,
 \end{equation}
with the (${\bf K}$-dependent) expectation values defined as 
space-time averages over the source function 
 \begin{equation}
 \label{4}
   \langle f(x) \rangle = {\int d^4x \, f(x) \, S(x,K) \over
                              \int d^4x \, S(x,K)} .
 \end{equation} 
Then the term $\delta S$ has vanishing zeroth, first and second order 
moments and thus contains only higher order information on sharp 
edges, wiggles, secondary peaks, etc. in the source. It was shown 
numerically \cite{WSH96} to have negligible influence on the half 
width of the correlation function and to contribute only weak, 
essentially unmeasurable structures in $C({\bf K},{\bf q})$ at large 
values of ${\bf q}$. Neglecting $\delta S$, the single-particle 
spectrum (\ref{5}) and the two-particle correlation function (\ref{1}) 
can be calculated analytically: 
 \begin{eqnarray}
 \label{10}
  E_K {dN \over d^3K} 
    &=& N({\bf K}) \, \det\left( (B^{-1})_{\mu\nu} ({\bf K})\right)
    \, S(\bar x({\bf K}),K)\, ,
 \\
 \label{11}
  C({\bf K},{\bf q}) 
    &=& 1 + \exp\left[ - q^\mu q^\nu 
            \langle \tilde x_\mu \tilde x_\nu \rangle ({\bf K}) 
                  \right] \, .
 \end{eqnarray}
The factor $\det(B^{-1}(K))$ in (\ref{10}) can be interpreted~\cite{CL96} 
as the generalized 4-volume of the emission region for particles with 
momentum $K$, $V_*^{(4)}({\bf K}) = \det\left(\langle \tilde x_\mu \tilde 
x_\nu \rangle \right)$. However, due to the $K$-dependent 
normalization factor $N(K)$, neither $V_*^{(4)}({\bf K})$ nor the 
point $\bar x^\mu({\bf K})$ of maximum emissivity at momentum ${\bf K}$ 
can be uniquely unfolded from the single-particle spectrum; the 
latter also drops out from the two-particle correlation function.
Only the (${\bf K}$-dependent) effective widths (``lengths of 
homogeneity" \cite{AS95,CSH95b}) $\langle \tilde x_\mu \tilde x_\nu 
\rangle ({\bf K})$ of the source of particles with momentum ${\bf K}$ 
are accessible by HBT interferometry. 

Furthermore, due to the on-shell constraint (\ref{2}), only 6 linear 
combinations of the variances $\langle \tilde{x}_\mu \tilde{x}_\nu 
\rangle({\bf K})$ are actually measurable \cite{CNH95}; in the case of 
azimuthal symmetry of the source around the beam axis, this number 
reduces to 4. To make contact with experimental correlation data, the 
redundant components must be eliminated from the exponent of the 
Gaussian in (\ref{11}). It is convenient to do this by using a 
cartesian coordinate system with $z$ along the beam axis and ${\bf K}$ 
lying in the $x$-$z$-plane. Customarily one labels the $z$-component 
of a 3-vector by $l$ (for {\em longitudinal}), the $x$-component by 
$o$ (for {\em outward}) and the $y$-component by $s$ (for {\em 
sideward}). Then from (\ref{2}) we see that $\beta_s=0$ such that 
 \begin{equation} 
 \label{12}
   q^0 = \beta_\perp q_o + \beta_l q_l
 \end{equation}
with $\beta_\perp = \vert {\bf K}_\perp \vert / K^0$ being 
(approximately) the velocity of the particle pair transverse to the 
beam direction while $\beta_l$ is its longitudinal component.

The standard form~\cite{CSH95b,CSH95a} for the parametrization of the 
correlation function is obtained by using (\ref{12}) to eliminate 
$q^0$ from Eq.~(\ref{11}). One obtains 
 \begin{equation}
     C({\bf K},{\bf q})
    = 1 + \exp\left[ -\sum_{i,j=s,o,l} R_{ij}^2({\bf K})\, q_i\, q_j 
              \right]
 \label{13}
 \end{equation}
where the 6 HBT radius parameters $R_{ij}$ are defined in terms of 
the following variances of the source function~\cite{CSH95a,CSH95b}:
 \begin{equation}   
   R_{ij}^2({\bf K}) = 
   \langle (\tilde{x}_i-{\beta}_i\tilde{t})
           (\tilde{x}_j-{\beta}_j\tilde{t})\rangle \, ,
   \quad i,j = s,o,l \, .
 \label{14}
 \end{equation}
For an azimuthally symmetric collision region or an azimuthally 
symmetric sample of collision events, $C({\bf K}, {\bf q})$
is symmetric with respect to $q_s \to -q_s$ \cite{CNH95}. Then
$R_{os}^2 = R_{sl}^2 = 0$ and
 \widetext
 \begin{equation}
    C({\bf K},{\bf q})
    = 1 + \exp\left[ - R_s^2({\bf K}) q_s^2 - R_o^2({\bf K}) q_o^2
                     - R_l^2({\bf K}) q_l^2 - 2 R_{ol}^2({\bf K}) q_o q_l
              \right] \, ,
 \label{15}
 \end{equation}
 \narrowtext
with
 \begin{mathletters}
 \label{16}
 \begin{eqnarray}   
   R_s^2({\bf K}) &=& \langle \tilde{y}^2 \rangle \, ,
 \label{16a}\\
   R_o^2({\bf K}) &=& 
   \langle (\tilde{x} - \beta_\perp \tilde t)^2 \rangle \, ,
 \label{16b}\\
   R_l^2({\bf K}) &=& 
   \langle (\tilde{z} - \beta_l \tilde t)^2 \rangle \, ,
 \label{16c}\\
   R_{ol}^2({\bf K}) &=& 
   \langle (\tilde{x} - \beta_\perp \tilde t)
           (\tilde{z} - \beta_l \tilde t) \rangle \, .
 \label{16d} 
 \end{eqnarray}
 \end{mathletters}
Clearly these HBT radius parameters mix spatial and temporal 
information on the source in a non-trivial way. Their interpretation 
in various reference systems, in particular the meaning of the 
generally non-vanishing cross-term $R_{ol}^2$, was extensively 
discussed in Refs.~\cite{CSH95b,CSH95a,WSH96,CNH95}, by analyzing 
these expressions analytically for a large class of (azimuthally 
symmetric) model source functions and comparing with the numerically 
calculated correlation function (\ref{1}). An important observation 
resulting from these studies is that the difference 
 \begin{equation}
 \label{17}
   R_{\rm diff}^2 \equiv  R_o^2 - R_s^2 =
   \beta_\perp^2 \langle \tilde t^2 \rangle - 2 \beta_\perp \langle
   \tilde{x} \tilde t\rangle + (\langle \tilde x^2 \rangle -
   \langle \tilde y^2 \rangle)
 \end{equation}
is generally dominated by the first term on the r.h.s. and thus 
provides access to the lifetime $\Delta t = \sqrt{\langle t^2 \rangle 
- \langle t \rangle^2}$ of the source \cite{CP91} (more exactly: the 
duration of the particle emission process). However, in heavy-ion 
collisions, due to rapid expansion of the source one would not expect
$\langle \tilde t^2 \rangle$ to be generically much larger than
either $\langle \tilde x^2 \rangle$ or  $\langle \tilde y^2 \rangle$; 
in the situations investigated so far (e.g. \cite{WSH96}) it comes 
out an order of magnitude smaller.
In the standard fit one is not sensitive to small values of $\Delta t$
since Eq.~(\ref{17}) then involves a small difference of two large
numbers, each associated with standard experimental errors. The
factor $\beta_\perp^2 \leq 1$ in front of $\langle \tilde t^2 \rangle$
further complicates its extraction, in particular at low $K_\perp$ 
where $\Delta t({\bf K})$ is usually largest (see below). Indeed, 
published experimental results \cite{NA35,NA44} so far show no 
positive evidence for a finite duration of the particle emission 
process, in contradiction to all physical intuition.

We will show here that a generalization to azimuthally symmetric 
systems \cite{CNH95,P83} of the Yano-Koonin parametrization for a 
moving source \cite{YK78} circumvents this problem \cite{fn1}. This 
Yano-Koonin-Podgoretskii (YKP) form is based on an elimination in 
Eq.~(\ref{11}) of $q_o$ and $q_s$ in terms of $q_{\perp} = \sqrt{q_o^2 
+ q_s^2}$, $q^0$, and $q_3$, using Eq.~(\ref{12}): 
 \widetext
 \begin{equation}
 \label{18}
   C({\bf K},{\bf q}) =
       1 +  \exp\left[ - R_\perp^2({\bf K})\, q_{\perp}^2 
                       - R_\parallel^2({\bf K}) \left( q_l^2 - (q^0)^2 \right)
                       - \left( R_0^2({\bf K}) + R_\parallel^2({\bf K})\right)
                         \left(q\cdot U({\bf K})\right)^2
                \right]  ,
 \end{equation}
 \narrowtext
where $U({\bf K})$ is a ($K$-dependent) 4-velocity with only a 
longitudinal spatial component:
 \begin{equation}
 \label{19}
   U({\bf K}) = \gamma({\bf K}) \left(1, 0, 0, v({\bf K}) \right) ,
   \ \ \text{with} \ \
   \gamma = {1\over \sqrt{1 - v^2}}\, .
 \end{equation}
This parametrization has the advantage that the YKP parameters 
$R_\perp^2({\bf K})$, $R_0^2({\bf K})$, and $R_\parallel^2({\bf K})$ 
extracted from such a fit do not depend on the longitudinal velocity 
of the observer system in which the correlation function is measured; 
they are invariant under longitudinal boosts. Their physical 
interpretation is easiest in terms of coordinates measured in the 
frame where $v({\bf K})$ vanishes. There they are given by 
\cite{CNH95} 
 \begin{mathletters}
 \label{20}
 \begin{eqnarray}   
 \FL
   R_\perp^2({\bf K}) &=& R_s^2({\bf K}) = \langle \tilde{y}^2 \rangle \, ,
 \label{20a} \\
   R_\parallel^2({\bf K}) &=& 
   \left\langle \left( \tilde z - {\beta_l\over\beta_\perp} \tilde x
                \right)^2 \right \rangle   
     - {\beta_l^2\over\beta_\perp^2} \langle \tilde y^2 \rangle 
     \approx \langle \tilde z^2 \rangle \, ,
 \label{20b} \\
   R_0^2({\bf K}) &=& 
   \left\langle \left( \tilde t - {1\over\beta_\perp} \tilde x
                \right)^2 \right \rangle 
    - {1\over\beta_\perp^2} \langle \tilde y^2 \rangle 
    \approx \langle \tilde t^2 \rangle \, ,
 \label{20c}
 \end{eqnarray}
 \end{mathletters}
where in the last two expressions the approximation consists of 
dropping terms which were found in \cite{CNH95} to be generically 
small (an extensive and quantitative discussion of this point will 
follow elsewhere \cite{WTWH96}). The first expression (\ref{20a}) 
remains true in any longitudinally boosted frame, and we will 
therefore now concentrate on the other three YKP parameters.  

Eq.~(\ref{20c}) shows that the YKP parameter $R_0({\bf K})$ 
essentially measures the time duration $\Delta t({\bf K})$ during 
which particles of momentum ${\bf K}$ are emitted, in the frame were 
the YKP velocity $v({\bf K})=0$. The crucial point here is that the
smallness of the difference $\langle \tilde x^2 - \tilde y^2 \rangle$ 
is already accounted for directly by the fit, and no potentially small
prefactor $\beta_\perp^2$ occurs. This means that the extraction
of $\Delta t({\bf K})$ from the YKP-parameter $R_0({\bf K})$ is much 
more direct and subject to less statistical uncertainties than in the
standard fit. Clearly, this point is only true and our suggestive
simple spatio-temporal interpretation of the YKP parameters is only 
valid as long as the approximations in (\ref{20}) are justified. For 
realistic emission functions they are as we shall show below.

Eqs.~(\ref{20}) were written down \cite{CNH95} in the special frame 
where $v({\bf K})=0$ which we call {\em Yano-Koonin (YK) frame}.
In \cite{CNH95} it was shown that 
for a large class of models this frame essentially coincides with the 
longitudinal rest frame of the fluid cell around the point $\bar 
x({\bf K})$ of maximum emissivity at momentum ${\bf K}$ (i.e. the {\em 
Longitudinal Saddle Point System} LSPS \cite{CL96}). This was true 
also for sources which are not longitudinally boost-invariant and for 
which the LSPS and the LCMS (the {\em Longitudinally CoMoving System} 
in which the pion pair has $\beta_l=0$ \cite{CP91}) do not coincide.  

We now give model independent expressions, similar to Eqs.~(\ref{14}), 
(\ref{16}) and (\ref{17}), for the YKP fit parameters in an arbitrary 
observer frame. They are again given in terms of second order 
moments of the source function $S(x,K)$ and thus calculable by simple 
quadrature. The expression for $v({\bf K})$ can then easily be used 
to establish, analytically and numerically, the relationship between 
the YK frame and the various other frames mentioned above. 

We introduce the following notational shorthands:
 \begin{mathletters}
 \label{21}
 \begin{eqnarray}
 \label{21a}
   A &=& \left\langle \left( \tilde t  
         - {\tilde \xi\over \beta_\perp} \right)^2 \right\rangle \, ,
 \\
 \label{21b}
   B &=&  \left\langle \left( \tilde z
         - {\beta_l\over \beta_\perp} \tilde \xi \right)^2 \right\rangle 
   \, ,
 \\
 \label{21c}
   C &=& \left\langle \left( \tilde t - {\tilde \xi\over \beta_\perp} \right)
                      \left( \tilde z - {\beta_l\over \beta_\perp} 
                             \tilde \xi \right) \right\rangle \, ,
 \end{eqnarray}
 \end{mathletters}
where $\tilde \xi \equiv \tilde x + i \tilde y$ and $\langle \tilde 
y\rangle = \langle \tilde x \tilde y \rangle = 0$ for azimuthally 
symmetric sources such that $\langle \tilde \xi^2 \rangle = \langle 
\tilde x^2 - \tilde y^2 \rangle$. In terms of these expressions one 
finds 
 \begin{mathletters}
 \label{22}
 \begin{eqnarray}
 \label{22a}
   v &=& {A+B\over 2C} \left( 1 - \sqrt{1 - \left({2C\over A+B}\right)^2}
                       \right) \, ,
 \\
 \label{22b}
   R_\parallel^2 &=& {1\over 2} \left( \sqrt{(A+B)^2 - 4C^2} - A + B
                        \right) = B{-}v C,
 \\
 \label{22c}
   R_0^2 &=& {1\over 2} \left( \sqrt{(A+B)^2 - 4C^2} + A - B 
                        \right) = A{-}v C , 
 \end{eqnarray}
 \end{mathletters}
The Yano-Koonin velocity $v$ is zero in the frame where the 
expression (\ref{21c}) for $C$ vanishes \cite{CNH95}; this fixes also 
the sign in front of the square root in (\ref{22a}). For small values 
of $C$ the Yano-Koonin velocity is given approximately by 
 \begin{equation}
 \label{23}
   v \approx {C\over A+B} 
     \approx {\langle \tilde z \tilde t \rangle \over
              \langle \tilde t^2 \rangle + \langle \tilde z^2 \rangle} 
 \, , 
 \end{equation}
where in the second approximation we again neglected generically small 
terms \cite{CNH95} proportional to $\langle \tilde z \tilde x\rangle$, 
$\langle \tilde x \tilde t \rangle$, and $\langle \tilde x^2 - \tilde 
y^2 \rangle$. The accuracy of the approximate expression (\ref{23}) 
for $v({\bf K})$ was tested numerically and found to be excellent in 
the situations discussed below. In the same limit the expressions for 
$R_0^2$ and $R_\parallel^2$ simplify to $R_0^2 \approx A$ and 
$R_\parallel^2 \approx B$, in agreement with (\ref{20}).  

It is instructive to compare the standard and YKP forms, 
Eqs.~(\ref{15}) and (\ref{18}), for the two-particle correlation 
function. One finds Eq.~(\ref{20a}) and 
 \begin{mathletters}
 \label{24}
 \begin{eqnarray}
 \label{24a}
   R_{\rm diff}^2 &=& R_o^2 - R_s^2 = \beta_\perp^2 \gamma^2 
             \left( R_0^2 + v^2 R_\parallel^2 \right) 
 \\
 \label{24b}
   R_l^2 &=& \left( 1 - \beta_l^2 \right) R_\parallel^2 
             + \gamma^2 \left( \beta_l-v \right)^2
             \left( R_0^2 + R_\parallel^2 \right)
 \\
 \label{24c}
   R_{ol}^2 &=& \beta_\perp \left( -\beta_l R_\parallel^2 
             + \gamma^2 \left( \beta_l-v \right)^2
             \left( R_0^2 + R_\parallel^2 \right) \right)
 \end{eqnarray}
 \end{mathletters}
To invert this set of equations we calculate (cf. Eqs.~(\ref{21}))
 \begin{mathletters}
 \label{25}
 \begin{eqnarray}
   A &=& {1 \over \beta_\perp^2} R_{\rm diff}^2 , 
 \label{25a} \\
   B &=& R_l^2 - {2\beta_l \over \beta_\perp} R_{ol}^2
         + {\beta_l^2 \over \beta_\perp^2} R_{\rm diff}^2  ,
 \label{25b} \\
   C &=& - {1\over \beta_\perp} R_{ol}^2 
         + {\beta_l\over \beta_\perp^2} R_{\rm diff}^2 .
 \label{25c}
 \end{eqnarray}
 \end{mathletters}
Inserting this into Eqs.~(\ref{22}) gives very cumbersome expressions 
which provide little physical insight. Thus, while the standard HBT 
radii are easily obtained from the YKP parameters via (\ref{24}), the 
converse is not true. This indicates that the YKP parameters are more 
``physical" than the standard HBT radii. Nevertheless, the relations 
(\ref{24}) provide a powerful consistency check on the experimental 
fitting procedure of the correlation function, of similar value as the 
relation \cite{CNH95,WSH96} $\lim_{K_\perp \to 0} (R_o({\bf K}) - 
R_s({\bf K})) = 0$ which results from azimuthal symmetry.  

We now discuss numerically the dependence of the YKP parameters 
on the pair momentum ${\bf K}$. For our study we use the model of 
Ref.~\cite{CNH95} for a finite expanding thermalized source
 \begin{eqnarray}
 \label{3.15}
  && S(x,K) = {M_\perp \cosh(\eta-Y) \over
            (2\pi)^3 \sqrt{2\pi(\Delta \tau)^2}}
 \\
 \nonumber
  && \quad \times \exp \left[- {K \cdot u(x) \over T}
                       - {(\tau-\tau_0)^2 \over 2(\Delta \tau)^2}
                       - {r^2 \over 2 R^2} 
                       - {{(\eta- \eta_0)}^2 \over 2 (\Delta \eta)^2}
           \right] .
 \end{eqnarray}
Here $r = \sqrt{x^2+y^2}$, the spacetime rapidity $\eta = {1 \over 2} 
\ln[(t+z)/(t-z)]$ and the longitudinal proper time $\tau= \sqrt{t^2-
z^2}$ parametrize the spacetime coordinates $x^\mu$, with measure 
$d^4x = \tau\, d\tau\, d\eta\, r\, dr\, d\phi$.  $Y = {1\over 2} 
\ln[(1+\beta_l)/(1-\beta_l)]$ and $M_\perp = \sqrt{m^2 + K_\perp^2}$ 
parametrize the longitudinal and transverse components of the pair 
momentum ${\bf K}$. $T$ is the freeze-out temperature, $R$ is the 
transverse geometric (Gaussian) radius of the source, $\tau_0$ its 
average freeze-out proper time, $\Delta \tau$ the mean proper time 
duration of particle emission, and $\Delta \eta$ parametrizes 
\cite{CSH95b} the finite longitudinal extension of the source. The 
expansion flow velocity $u^\mu(x)$ is parametrized as 
 \begin{equation}
 \label{26}
   u^\mu(x) = \left( \cosh \eta \cosh \eta_t(r), \,
                     \sinh \eta_t(r)\, {\bf e}_r,  \,
                     \sinh \eta \cosh \eta_t(r) \right) ,
 \end{equation}
with a boost-invariant longitudinal flow rapidity $\eta_l = \eta$ and 
a linear transverse flow rapidity profile
 \begin{equation}
 \label{27}
  \eta_t(r) = \eta_f \left( {r \over R} \right)\, .
 \end{equation} 
$\eta_f$ scales the strength of the transverse flow. Other 
possible features of the source, like spatial and temporal gradients 
of the freeze-out temperature \cite{CL96}, other freeze-out 
hypersurfaces or different flow profiles, will be discussed elsewhere.

For the numerical calculations in this letter we have selected one 
fixed set of source parameters: $R=3$ fm, $\tau_0 = 3$ fm/$c$, $\Delta 
\tau = 1$ fm/$c$, $\Delta \eta = 1.2$, $T=140$ MeV. We study only pion 
correlations and set $m=m_\pi=139$ MeV/$c^2$. Results for different 
parameter sets as well as for kaon correlation functions will be 
presented in a longer paper \cite{WTWH96}.

In Fig.~\ref{F1} we show the relationship between the YK frame and the 
LCMS and LSPS. $Y$ is the pion pair rapidity (and thus the rapidity of 
the LCMS), $Y_{_{\rm YK}}(Y,K_\perp)$ the rapidity of the Yano-Koonin 
rest frame, and $Y_{_{\rm LSPS}}(Y,K_\perp)$ the rapidity of the 
longitudinal rest frame of the point $\bar x(Y,K_\perp)$ of maximum 
emissivity (all rapidities are measured relative to the CMS of the 
source). For pion pairs with large $K_\perp$ both the YK rest frame 
and the LSPS rapidities approach the LCMS rapidity $Y$, i.e. in this 
limit all the pions are emitted from a small region in the source 
which moves with the same longitudinal velocity as the pion pair. For 
small $K_\perp$ the YK frame is considerably slower than the LCMS, but 
faster than the LSPS. {\em The linear relationship between the rapidity 
$Y_{_{\rm YK}}$ of the Yano-Koonin frame and the pion pair rapidity 
$Y$ is a direct reflection of the boost-invariant longitudinal 
expansion flow.} Such a behaviour, and thus direct evidence for a 
strong longitudinal expansion of the source, was recently found 
experimentally in Mg+Mg collisions at 4.4 $A$ GeV/$c$ in Dubna \cite{GIBS96}. 

The difference between $Y_{_{\rm YK}}$ and $Y_{_{\rm LSPS}}$ is due to 
a longitudinal asymmetry of the source around the saddle point $\bar 
x(Y,K_\perp)$; if the source is $z$-symmetric around $\bar 
x(Y,K_\perp)$ the YK rest frame and the LSPS become identical 
\cite{WTWH96}. Both $Y_{_{\rm YK}}$ and $Y_{_{\rm LSPS}}$ exhibit only 
a very weak dependence on the transverse flow of the source; its 
origin will be discussed quantitatively in \cite{WTWH96}.  

In Fig.~\ref{F2} we show $R_0$ and $R_\parallel$ as a function of 
$K_\perp$ for pion pairs with momentum $Y=0$ and $Y=3$ in the CMS 
frame and compare these radii with the approximations $R_\parallel 
\approx \sqrt{\langle \tilde z^2 \rangle}$, $R_0 \approx \sqrt{\langle 
\tilde t^2 \rangle}$ given in Eqs.~(\ref{20}b,c). The approximation is 
seen to be {\em exact for vanishing transverse flow,} $\eta_f=0$ (as 
already pointed out in \cite{CNH95}). For $R_\parallel$ it remains 
rather accurate for all $K_\perp$-values even in the presence of
large transverse flow (Fig.~\ref{F2}d). The parameter $R_0$, on the 
other hand, is an accurate measure of $\Delta t$ only for small 
$K_\perp$ or sufficiently small transverse flow \cite{CNH95}. The 
difference between these two quantities arises from the terms
$-2\langle \tilde x \tilde t \rangle/\beta_\perp + \langle \tilde
x^2 - \tilde y^2 \rangle/\beta_\perp^2$ which were neglected in the
second equality of Eq.~(\ref{20c}). As seen in Figs.~\ref{F2}a,b, 
these terms can become a serious source of error in the determination
of $\Delta t({\bf K})$ (in our case an overestimate of up to 50\% 
in the most unfavorable case) if the transverse flow of the source 
is very large and not independently known such that it could be 
corrected for. However, it should be noted \cite{CNH95} that the 
contamination by these undesired terms is absent for pion pairs 
with small pair momentum ${\bf K}$, and that therefore the 
determination of $\Delta t$ from the YKP-parameter $R_0$
is particularly clean in the region where its extraction from the 
standard fit according to Eq.~(\ref{17}) is difficult due to the 
$\beta_\perp^2$-prefactor. Furthermore, the terms that contaminate
$R_0^2$ at large ${\bf K}$ {\em and} large transverse flow affect 
the extraction of $\Delta t$ from Eq.~(\ref{17}) in exactly the same
way. This problem can thus not be avoided be selecting either the 
standard or the YKP fitting procedure; by doing and comparing both, 
in particular also for heavier particles, it may be possible to 
estimate the amount of transverse flow and correct for it \cite{WTWH96}.
Here it should suffice to say that the associated relative error on 
$\Delta t$ is everywhere less than 25\% for transverse expansion 
velocities $\eta_f \leq 0.3$ which we believe to be realistic, and that 
it should decrease for more realistic larger values for the model 
parameter $\Delta \tau$ than the 1 fm/$c$ chosen in Fig.~\ref{F2}.

Both the longitudinal region of homogeneity $\sqrt{\langle \tilde z^2 
\rangle}$ and the effective lifetime $\sqrt{\langle \tilde t^2 
\rangle}$ of the source decrease for pion pairs with large momenta in 
the CMS of the source. Asymptotically the effective lifetime becomes 
equal to the model parameter $\Delta \tau = 1$ fm/$c$, but 
low-momentum pions see a much larger value. This is because for low 
pair momenta the longitudinal region of homogeneity $R_\parallel$ is 
large, and the correlation function receives also contributions from 
regions freezing out at later times $t = \sqrt{\tau_0^2 + z^2} \pm 
\Delta \tau$ along the surface of constant proper time $\tau_0$ ($z$ 
and $t$ measured in the YK rest frame). This is a generic effect which 
should also appear for different source models \cite{fn2}. The 
resulting strong variation of $\langle \tilde t^2 \rangle$ at small 
$K_\perp$ is again hard to extract from the standard fit because this 
region is suppressed by the factor $\beta_\perp^2$ in Eq.~(\ref{17}). 
Although for large transverse flow $R_0$ does no longer exactly trail 
the lifetime $\langle \tilde t^2 \rangle$, it clearly reflects this 
strong ${\bf K}$-dependence of the latter at small values of ${\bf K}$.  

To summarize, we have given model-independent expressions and a 
detailed physical interpretation of the fit parameters for a 
Yano-Koonin-Podgoretskii fit to the two-particle correlation function. 
We have also established a simple analytical relation between these 
parameters and the ``standard" HBT radius parameters which 
provides a powerful consistency check on the experimental fitting 
procedure. We clarified the relationship between the YK rest frame and 
the previously introduced LCMS and LSPS frames and argued that the YKP 
fit parameters provide the most intuitive characterization of the 
local geometric and dynamical space-time characteristics of the source.
An increase of the YK velocity with the pair rapidity signals 
longitudinal expansion of the source. We also pointed out a strong 
generic ${\bf K}$-dependence of the effective duration of particle 
emission which results mainly from the fast longitudinal expansion of 
the source, but is also modulated by transverse expansion. We hope 
that all these predictions will soon be checked experimentally in 
relativistic heavy ion collisions.  

This work was supported by grants from DAAD, DFG, NSFC, BMBF and GSI. 
We gratefully acknowledge discussions with H. Appelsh\"auser, 
S. Chapman, D. Ferenc, M. Ga\'zdzicki, and P. Seyboth.  

%%%%%%%%%%%%%%%%%%%%%%%%%%%%%%%%%%%%%%%%%%%%%%%%%%%%%%%%%%%%%%%%%%%%%%

\begin{figure}
\caption{
 (a) The rapidity of the YK frame as a function of the pion 
     pair rapidity $Y$ (both measured in the CMS frame of the source), for 
     various values of the transverse momentum $K_\perp$ of the pair and 
     two values for the transverse flow rapidity $\eta_f$.
 (b) Same as (a), but shown as a function of $K_\perp$ for different 
     values of $Y$. The curves for negative $Y$ are obtained by 
     reflection along the abscissa.  
 (c) The difference $Y_{_{\rm YK}} - Y_{_{\rm LSPS}}$ between the 
     rapidity of the YK frame and the longitudinal rest system of the 
     saddle point, plotted in the same way as (a). 
 (d) Same as (c), but shown as a function of $K_\perp$ for different 
     values of $Y$. 
}
\label{F1}
\end{figure}

\begin{figure}
\caption{
 (a) $R_0$ and $\protect \sqrt{\langle \tilde t^2 \rangle}$ 
     as a function of $\protect M_\perp$ for three values of 
     the transverse flow rapidity $\eta_f$, for pion pairs with 
     rapidity $Y=0$ in the source CMS frame. The lifetime $\protect 
     \sqrt{\langle \tilde t^2 \rangle}$ is evaluated in the YK rest 
     frame (which in this case coincides with the CMS frame).  
 (b) Same as (a), but for pions with rapidity $Y=3$ in the CMS frame. 
 (c) and (d): Same as (a) and (b), but for $R_\parallel$ and the 
     longitudinal length of homogeneity $\protect \sqrt{\langle \tilde 
     z^2 \rangle}$ in the YK rest frame. For $Y=0$, $R_\parallel$ and 
     $\protect \sqrt{\langle \tilde z^2 \rangle}$ agree exactly 
     because $\beta_l=0$ in the YK frame.
}
\label{F2} 
\end{figure} 


\begin{references}
\bibitem{fn2}
  One might argue that this effect should disappear if freeze-out 
  occurs at constant coordinate time $t_0$ rather than at constant 
  proper time $\tau_0$. However, this is then true only for pions 
  whose YK rest frame coincides with the system where $t_0$ is 
  constant. Pion pairs with different momenta will on average 
  originate from source regions with different YK velocities, and in 
  their respective YK rest frames the freeze-out surface will again 
  extend over a finite time interval which can be probed by 
  low-$K_\perp$ pions.
\end{references}
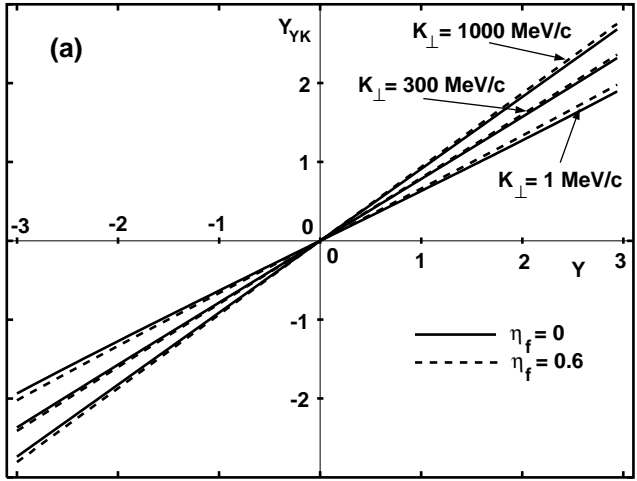
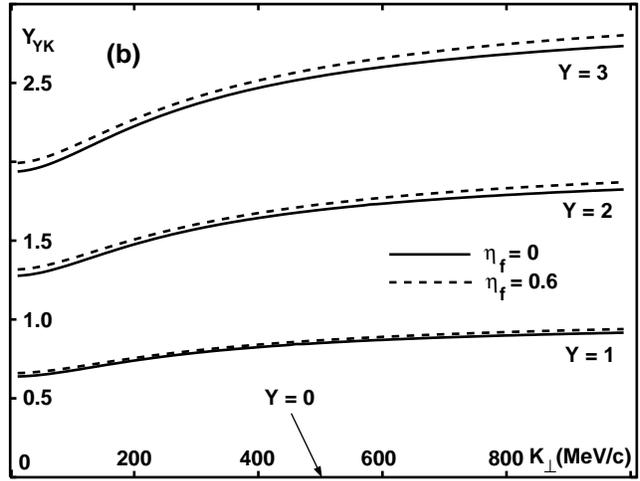
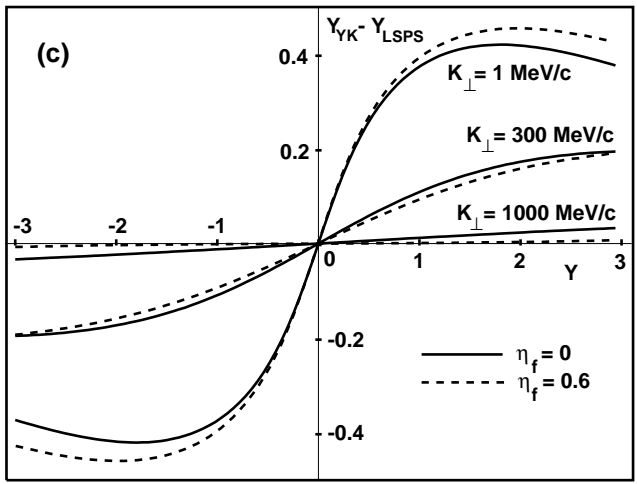
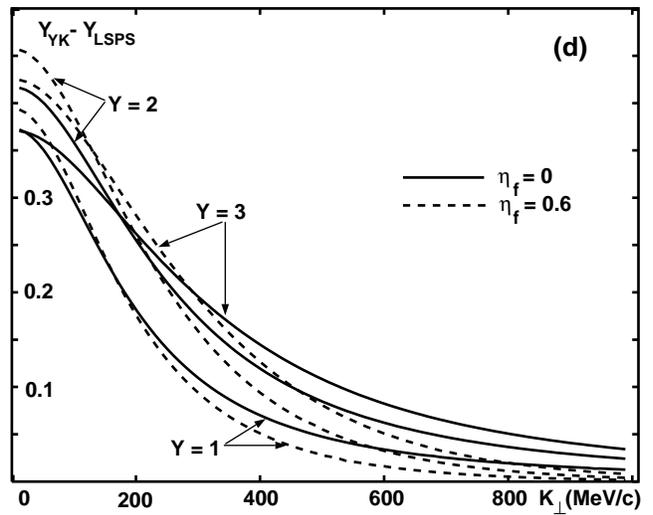
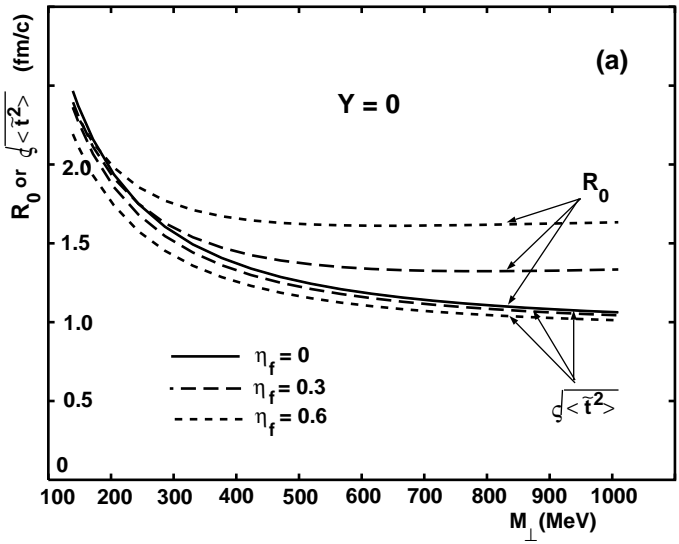
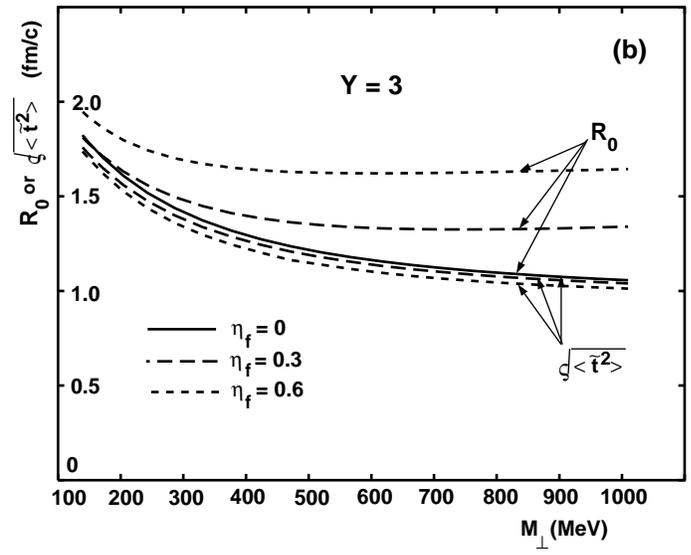
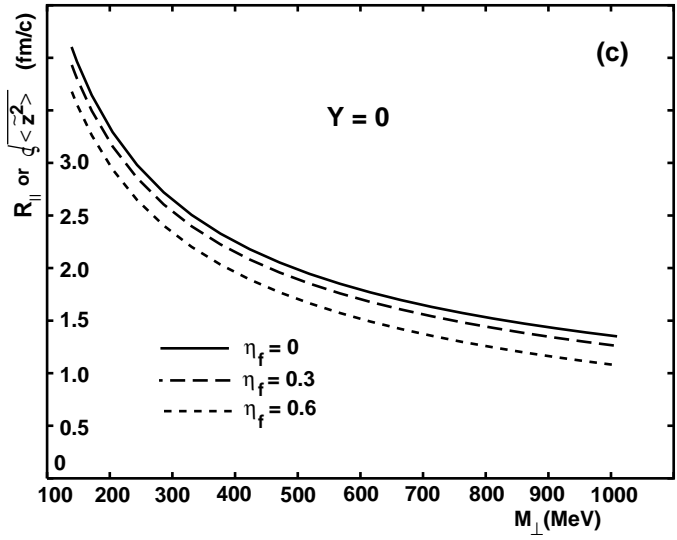
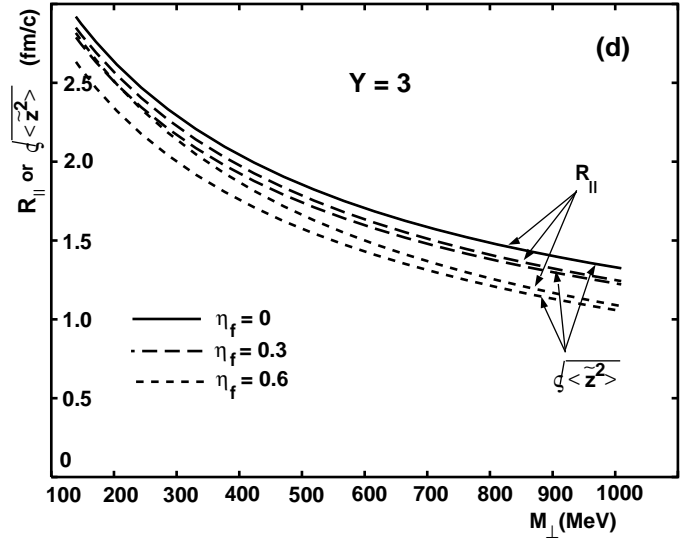
\end{document}